\documentclass{article}

\begin{document}
\newcommand{\nd}{\noindent}
\newcommand{\nl}{\newline}
\newcommand{\be}{\begin{equation}}
\newcommand{\ee}{\end{equation}}
\newcommand{\ben}{\begin{eqnarray}}
\newcommand{\een}{\end{eqnarray}}
\newcommand{\nn}{\nonumber \\}
\newcommand{\ii}{\'{\i}}
\newcommand{\pp}{\prime}
\newcommand{\expq}{e_q}
\newcommand{\lnq}{\ln_q}
\newcommand{\quno}{q-1}
\newcommand{\qunoinv}{\frac{1}{q-1}}
\newcommand{\tr}{{\mathrm{Tr}}}



\title{The mutual co-implication of thermodynamics' first and second laws}

\author{A. Plastino$^1$ and E. M. F. Curado$^2$  \\
$^1$ La Plata National University \\ Argentina's National Research Council (CONICET)\\
C. C. 727 - 1900 La Plata - Argentina \\ E-mail: plastino@fisica.unlp.edu.ar\\
$^2$ Centro Brasileiro de Pesquisas Fisicas (CBPF)\\ Rua Xavier Sigaud 150 - Urca - Rio de Janeiro - Brasil\\E-mail: evaldo@cbpf.br}

\maketitle

\begin{abstract}
 \nd In classical phenomenological thermodynamics the first and second laws can be regarded as independent statements. Statistical mechanics provides a microscopic substratum that explains thermodynamics in probabilistic terms via a microstate probability distribution $\{p_i\}$. We study here a hitherto unexplored  microscopic connection between the two laws. Given an information measure (or entropic form), each of the       two laws implies the other through the process $p_i \rightarrow p_i+dp_i$.
\vskip 3mm

PACS: 2.50.-r, 3.67.-a, 05.30.-d
\vskip 3mm
KEYWORDS: Thermodynamics, Microscopic probability distribution, First law, Second law.
 \end{abstract}

 \newpage


\section{Introduction}

\nd Macroscopically,
in classical phenomenological thermodynamics,
the first and second laws
can be regarded as independent statements.
In statistical mechanics an
underlying microscopic substratum
is added that is able to explain thermodynamics itself \cite{patria,reif,sakurai,katz}.

\nd Of this substratum, a microscopic
probability distribution (PD) that controls
the population of microstates is a basic ingredient \cite{patria}.
Changes that affect exclusively microstate-population
give rise to ``heat" \cite{reif,katz}.
How these changes are related to
energy-changes provides the essential content of the first law \cite{reif}.
\vskip 4mm
\nd In this effort we show that the above
mentioned  PD establishes a link between the first
and second laws of thermodynamics, according to the following scheme:
\begin{itemize}
\item Given an entropic form (or an information measure (IM)) $S$,
a mean energy $U$ and a temperature $T$,
\item and for any system described
by a microscopic probability distribution (PD) $\{p_i\}$,
 \item  assuming a heat transfer process via $p_i \rightarrow p_i+dp_i$,
 \item 1) if the  PD $\{p_i\}$ maximizes $S$ this entails  $dU=TdS$, and, alternatively,
  \item 2) if $dU=TdS$, this predetermines a unique PD that maximizes $S$.
 \end{itemize}
\nd Symbolically, given a specific  IM,
$$dU=TdS\,\,\Leftrightarrow\,\,{\rm MaxEnt\,\, prob.\,\, distr.\,\, \{p_i\}}.$$

\section{From the second to the first law}

\nd A quite general treatment is given in, for instance
\cite{m1,e1} (by no means an exhaustive list!). Here, for
completeness' sake, we provide a rudimentary sketch of the
pertinent arguments.

\nd One way to  be sure that one complies with the strictures of
the second law is to use MaxEnt \cite{katz}, i.e., maximize the
entropy $S$ with, say $M$, appropriate constraints. If the
pertinent microstates are denoted with the subindex $i$, and the
physical quantity $A_k,\,\,\,\,(k=1,\ldots,M)$ takes the value
$A_k(i)$ at the microstate $i$, then the constraints read

\be \label{s1} \langle A_k \rangle= \sum_i\,p_i\,A_k(i);\,\,\,\,(k=1,\ldots,M).\ee

\nd We will denote the Boltzmann constant by $k_B$ and assume that
$k=1$ corresponds to the energy $E$ with
 $(A_1(i)\equiv \epsilon_i)$, so that, in such a case the above expression specializes to

\be \label{s2} U\equiv \langle A_1 \rangle= \sum_i\,p_i\,\epsilon_i.\ee
\nd One should now maximize the ``Lagrangian" $\Phi$ \cite{katz}

\be \label{s3} \Phi= S/k_B-\alpha
 \sum_i \,p_i \,-
 \beta \sum_i \,p_i\,\epsilon_i\,-\,\sum_{k=2}^M\,\lambda_k\,\sum_i \,p_i\,A_k(i),   \ee
in order to obtain the actual distribution $\{p_i\}$ from the equation

\be \label{s4}  \delta_{p_i}\Phi=0. \ee However, since in this
paper we are interested just in the ``heat" part we shall not
consider the last term on the right-hand-side of (\ref{s3}). We
argue that, if $p_i$ changes to $p_i\,+\,dp_i$, because of
(\ref{s4}) we have

\be \label{s5} 0=dS/k_B\,-\beta \,dU,\ee which implies (note that,
because of normalization, $\sum_i\,\delta p_i=0$), with
$\beta=1/k_BT$, $T$ the temperature \cite{reif} \be \label{s6}
dU=TdS,\ee which concludes the argument \cite{m1} that MaxEnt
leads to the first law. \vskip 4mm \nd The whole procedures given
both in this and  in the forthcoming Section can be repeated
including also the {\it work} ``$\lambda_k\,A_k$"-terms of
(\ref{s3}), of course. Such an extension is straightforward and we
omit it for brevity's sake.

\section{From the first law to the second}
\nd Our {\it central goal is to traverse now the opposite
direction} as that of the preceding Section. We shall start the
present considerations by assuming that one deals with a rather
{\it general} information measure of the form

\be \label{1} S= k\sum_i\,p_i\,f(p_i), \ee where, for simplicity's
sake, Boltzmann's constant is denoted now just by $k$. The sum
runs over a set of quantum numbers, collectively denoted by $i$
(characterizing levels of energy $\epsilon_i$), that specify an
appropriate basis in Hilbert's space and $\mathcal{P}=\{p_i \}$ is
an (as yet unknown) un-normalized probability distribution such
that

\be \label{2} \sum_i\,p_i= constant.\ee
Finally, $f$ is an arbitrary  smooth function of the $p_i$.
Further, we assume that mean
values of quantities $A$
that take the value $A_i$ with probability $p_i$ are evaluated according to

\be \label{3} \langle A \rangle=
\sum_i\,A_i\,g(p_i), \ee with $g$ another arbitrary
smooth function of the $p_i$. In particular, the mean energy $U$ is given by
\be \label{4} U= \sum_i\,\epsilon_i\,g(p_i).\ee

\nd Assume now that the set $\mathcal{P}$ changes in the fashion
\be \label{5} p_i \rightarrow p_i \,+\, dp_i,\,
{\rm with}\,\,\sum_i\,dp_i=0\,\,\,({\rm on\,account\,of\,(\ref{2})}),\ee
which in turn generates corresponding changes $dS$ and $dU$
in, respectively, $S$ and $U$.
We are talking just about level-population changes, i.e., heat.
We want then to make sure that the heat part of
thermodynamics' first law is obeyed,
so that we impose the condition that, in the above described circumstances,
\be \label{6} dU-TdS=0, \ee
with $T$ the temperature.
As a consequence of (\ref{6}),
a little algebra yields, up to first order in the $dp_i$, the condition

\be \label{7} \epsilon_i\,g'(p_i)-kT\left[f(p_i)\,+\,p_i\,f'(p_i)
\right]=0, \ee
where the primes indicate derivative with respect
to $p_i$. Eq. (\ref{7}) should hopefully yield one and just one
expression for the $p_i$. We proceed to show that this is indeed
the case by examining below several important situations.

 \section{Shannon's entropy}

 \nd Here we take
 \be \label{8} f(p_i)= -\ln{(p_i)},\,\,\,\,{\rm and}\,\,\, g(p_i)=p_i. \ee
 In these circumstances, Eq. (\ref{7}) becomes

 \be \label{9} -\epsilon_i=kT[\ln{(p_i)}+1], \ee
 which immediately yields (remember (\ref{5}))

 \be \label{10} p_i=\frac{1}{e}\exp{(-\epsilon_i/kT)},\ee
 that after normalization yields the canonical
 Boltzmann distribution (BD).
 We conclude that this distribution is the only one that
 guarantees obedience to the first law for Shannon's
 information measure. A posteriori, one ascertains that the
 BD maximizes entropy as well, with $U$ as a constraint,
 which establishes a link with the second law.

 \section{Tsallis measure with linear constraints}

 \nd We have now, for any real number $q$ \cite{tsallis,Brasil,PP},

  \be \label{11} f(p_i)= \,\frac{(1\,-\,p_i^{q-1})}{q-1},
  \,\,\,\,{\rm and}\,\,\, g(p_i)=p_i, \ee
  so that $f'(p_i)=-\,p_i^{q-2}$ and Eq. (\ref{7}) becomes, with $\beta=(1/kT)$,

 \be \label{12}  q \,p_i^{q-1}= 1-(q-1)\,\beta\, \epsilon_i , \ee
which after normalization yields Tsallis' celebrated 1988 distribution

  \ben \label{13} p_i&=& Z_q^{-1}
  \left[1-(q-1)\,\beta\, \epsilon_i \right]^{1/(q-1)} \cr
   Z_q&=& \sum_i\,\left[1-(q-1)\,\beta\, \epsilon_i \right]^{1/(q-1)}. \een

  \section{Tsallis measure with non-linear constraints, un-normalized}

 \nd The information measure is still the one  built up
 with the function $f(p_i)$ of (\ref{11}),
 but we use now the so-called Curado-Tsallis 1991
 constraints \cite{CT} that arise if one uses

 \be \label{14} g(p_i)=p_i^q\,\,\Rightarrow\,
 \,g'(p_i)=q\,p_i^{q-1}. \ee Eq. (\ref{7}) leads to

 \be \label{15} \,p_i=(\frac{1}{q})^{1/(q-1)}
  \left[1\,-\,(1-q)\,\beta  \epsilon_i\right]^{1/(1-q)}, \ee
  and, after normalization, one is led to the Curado-Tsallis distribution \cite{CT}

  \ben \label{17} p_i&=& (Z_q)^{-1}
  \,\left[1-(1-q)\,\beta\, \epsilon_i \right]^{1/(1-q)} \cr
   Z_q&=& \sum_i\,\left[1-(1-q)\,\beta\, \epsilon_i \right]^{1/(1-q)}. \een

  \section{Tsallis measure with non-linear constraints, normalized}

 \nd This is the standard treatment nowadays \cite{tsallis}.
 It was proposed in \cite{TMP}. One has

 \be \label{100}
 g(p_i)= \frac{p_i^q}{w_q};\,\,\,\,\,w_q
 =\sum_i\,p_i^q;\,\,\,\,\,U_q=\sum_i\,g(p_i)\,\epsilon_i,\ee
 which entails
 \be \label{101}
 g'(p_i)= \frac{qp_i^{q-1}}{w_q}\,[1\,-\,\frac{p_i^q}{w_q}].  \ee
 This is to be inserted into (\ref{7}) and one finds

 \be \label{102} (1-q)\,\beta\epsilon_i\,g'(p_i)= qp_i^{q-1}\,-\,1 \ee i.e.,

 \be \label{103}  qp_i^{q-1} = [1+
  \frac{(1-q)\beta \,qp_i^{q-1}\,\epsilon_i}{w_q}\,(1\,-\,\frac{p_i^q}{w_q})].  \ee
 Now, we see that $qp_i^{q-1}$  is the common factor of the quantity $\mathcal{C}$
 \be \mathcal{C}= 1\,-\,\frac{(1-q)\beta \epsilon_i}{w_q}\,(1\,-\,\frac{p_i^q}{w_q})
\,\,\, \Rightarrow\,\,\,qp_i^{q-1}\mathcal{C}= 1,\ee so that
 \be \label{300} \frac{1}{
qp_i^{q-1}}\,=\,\mathcal{C}=\,\left[1-(1-q)\,\beta\,\frac{\epsilon_i}{w_q}\,(1\,-\,\frac{p_i^q}{w_q})
\right],\ee

   \nd If in the equality above we sum  over the running index $i$ we get

  \ben \label{104} \mathcal{S}&=&
  \sum_i\, \frac{1}{qp_i^{q-1}}\,-\,\sum_i
  \,\left(\frac{(1-q)\beta }{w_q}\,
   \epsilon_i\,[1\,-\,\frac{p_i^q}{w_q}]\right)\cr &\equiv&
   \sum_i\, \frac{1}{qp_i^{q-1}}-\mathcal{S}_1+\mathcal{S}_2 =0,
   \een
   with
   \ben \mathcal{S}_1 &=& \sum_i
  \,\frac{(1-q)\beta }{w_q}\,
   \epsilon_i   \cr \mathcal{S}_2&=&\,\sum_i
  \,\frac{(1-q)\beta }{w_q}\,
   \epsilon_i\,\frac{p_i^q}{w_q},\een
Consider now $\mathcal{S}_2$. It acquires the appearance \be
\mathcal{S}_2 =
  \frac{(1-q)\beta }{w_q}\,\sum_i\,
   \epsilon_i\,\frac{p_i^q}{w_q}=\frac{(1-q)\beta }{w_q}\,U_q, \label{44} \ee
i.e., it is a {\it constant} independent of $i$. Assume that, for
the system one is interested in, the spectrum consists of
$\mathcal{N}$ nondegenerate energy levels $i$. One can then rearrange things in
(\ref{44})  so as to add together $\mathcal{N}$ terms $U_q$ in
$\mathcal{S}_2$, using the obvious trick
$\mathcal{S}_2=\mathcal{S}_2(\mathcal{N}/\mathcal{N})$, so that we
have \be -\mathcal{S}_1+\mathcal{S}_2 = \sum_i^{\mathcal{N}}
  \,\left(\frac{(1-q)\beta }{w_q}\,
   [\epsilon_i\,-\,\frac{U_q}{\mathcal{N}}]\right).\ee
  Tsallis et al. \cite{TMP} argue
at this point that one is, of course,  free to shift the energy
scale so as to add a fixed amount
$W=U_q\frac{\mathcal{N}-1}{\mathcal{N}}$ to each $\epsilon_i$,
according to the  Mach's dictum: {\it no absolute origins}. Since
the origin of the energy spectrum can always be freely chosen, one
can legitimately assume then the {\it uniform} energy-shift

$$\epsilon_i \mapsto \varepsilon_i;\,\,\,\varepsilon_i=\epsilon_i
+ U_q\frac{\mathcal{N}-1}{\mathcal{N}}.$$ \nd  This enables
Tsallis et al. to write \cite{TMP} \be
-\mathcal{S}_1+\mathcal{S}_2 = \sum_i^{\mathcal{N}}
  \,\left(\frac{(1-q)\beta }{w_q}\,
   [\varepsilon_i\,-\,U_q]\right) \, ,  \ee 
 and argue that $\mathcal{S}$ clearly vanishes 
 if, in the first line of (\ref{104}), that now reads

\be \label{1044} \mathcal{S}=
  \sum_i\, \frac{1}{qp_i^{q-1}}\,-\,\sum_i
  \,\left(\frac{(1-q)\beta }{w_q}\,
   [\varepsilon_i\,-U_q]\right), \ee
   each $i-$term vanishes by itself (with
  $\epsilon$ replaced by $\varepsilon$).
  This  prompts one to write the pertinent,
  properly normalized probability distribution in the TMP fashion \cite{TMP}

  \ben \label{105} p_i &=&
  Z_q^{-1}\,[1\,-\, \frac{(1-q)\beta }{w_q}
  \,(\varepsilon_i\,-\,U_q)]^{1/(1-q)}
   \cr     Z_q&=&\sum_i\,
   \left[1\,-\, \frac{(1-q)\beta }{w_q}\,(\varepsilon_i\,-\,U_q)\right]^{1/(1-q)}.\een

  \section{Exponential entropic form}

  \nd This is given in \cite{e1,e2} and also used in \cite{NC}. One has

  \be \label{18} f(p_i)= \frac{1-\exp{(-bp_i)}}{p_i} - S_0, \ee
  \nd where $b$ is a positive constant and $S_0 = 1-\exp(-b)$, together with

  \be \label{19} g(p_i)= \frac{1-e^{-bp_i}}{S_0}\,\,
  \,\Rightarrow \,\,\,g'(p_i)=\frac{be^{-bp_i}}{S_0},\ee
  which, inserted into (\ref{7}), after a little algebra, leads to
  \be \label{20} p_i=\frac{1}{b}\,
  \left[\ln{\frac{b}{S_0}}\,+\,\ln{(1\,- \,\frac{\beta\epsilon_i}{S_0})}\right].
\ee
which, after normalization, gives the correct answer \cite{e2}.

    \section{Conclusion}

    \nd We have endeavored to show in this communication that,
    from a microscopic perspective,
    the first and second law of thermodynamics
    co-imply themselves in reciprocal fashion, that is

\begin{itemize}
\item assuming entropy is maximum one immediately derives the first law, and
\item if you assume the
validity of the first law and
an information measure,
this predetermines a probability distribution that maximizes entropy.
\end{itemize}
   \nd The first item has been known for some time
   (see, for instance, \cite{m1,e1}).
   As far as we know, the present is the first
   instance in which the second item has received detailed discussion.

  \end{document}